\documentclass[sigconf]{acmart}
\usepackage{graphicx}
\usepackage{multirow}
\usepackage{amsmath}
\usepackage{url}
\usepackage{pifont}
\usepackage{subcaption}
\usepackage{rotating}
\usepackage{enumitem}
\setlist[itemize]{noitemsep, topsep=0pt, nosep, leftmargin=*}
\setlist[enumerate]{noitemsep, topsep=0pt, nosep, leftmargin=23pt}
\DeclareMathOperator*{\argmin}{argmin}
\newcommand{\cmark}{\ding{51}}
\newcommand{\xmark}{\ding{55}}
\newcommand{\C}[1]\null
\newcommand{\R}[1]{\begin{sideways}{#1}\end{sideways}}

\copyrightyear{2024}
\acmYear{2024}
\setcopyright{acmlicensed}\acmConference[RecSys '24]{18th ACM Conference on Recommender Systems}{October 14--18, 2024}{Bari, Italy}
\acmBooktitle{18th ACM Conference on Recommender Systems (RecSys '24), October 14--18, 2024, Bari, Italy}
\acmDOI{10.1145/3640457.3688073}
\acmISBN{979-8-4007-0505-2/24/10}

\begin{document}
\title{Revisiting BPR: A Replicability Study of a Common Recommender System Baseline}

\author{Aleksandr Milogradskii}
\email{alex.milogradsky@gmail.com}
\affiliation{
  \institution{National Research University Higher School of Economics, T-Bank}
  \country{Russia}
}

\author{Oleg Lashinin}
\email{fotol764@gmail.com}
\affiliation{
  \institution{Moscow Institute of Physics and Technology, T-Bank}
  \country{Russia}
}

\author{Alexander P}
\email{alexander.p.89@yandex.ru}
\affiliation{
  \institution{Independent Researcher}
  \country{Russia}
}

\author{Marina Ananyeva}
\email{ananyeva.me@gmail.com}
\affiliation{
  \institution{National Research University Higher School of Economics, T-Bank}
  \country{Russia}
}

\author{Sergey Kolesnikov}
\email{scitator@gmail.com}
\affiliation{
  \institution{T-Bank}
  \country{Russia}
}

\begin{abstract}
Bayesian Personalized Ranking (BPR), a collaborative filtering approach based on matrix factorization, frequently serves as a benchmark for recommender systems research. However, numerous studies often overlook the nuances of BPR implementation, claiming that it performs worse than newly proposed methods across various tasks. In this paper, we thoroughly examine the features of the BPR model, indicating their impact on its performance, and investigate open-source BPR implementations. Our analysis reveals inconsistencies between these implementations and the original BPR paper, leading to a significant decrease in performance of up to 50\% for specific implementations. Furthermore, through extensive experiments on real-world datasets under modern evaluation settings, we demonstrate that with proper tuning of its hyperparameters, the BPR model can achieve performance levels close to state-of-the-art methods on the top-n recommendation tasks and even outperform them on specific datasets. Specifically, on the Million Song Dataset, the BPR model with hyperparameters tuning statistically significantly outperforms Mult-VAE by 10\% in NDCG@100 with binary relevance function. 
\end{abstract}

\maketitle
\section{Introduction}

\looseness -1 The issue of information overload, coupled with the rise of online services, has created a growing need for recommender systems \cite{maes1995agents}. These systems have attracted significant interest from both academic \cite{jannach2021survey, wang2021survey, afsar2022reinforcement, wu2022graph} and industrial researchers \cite{covington2016deep, liang2018variational, spotifyrec}, who have studied various methods associated with personalization.

\looseness -1 Traditionally, recommender systems have relied on either explicit \cite{bennett2007netflix} or implicit \cite{hu2008collaborative} feedback. Explicit actions, such as user ratings, are relatively limited. In contrast, implicit feedback, such as views, clicks, and purchases, is vast in number, making it easier to build recommender systems based on these logs \cite{BPR}. However, these models are less confident \cite{hu2008collaborative}.

\begin{figure}[t]
    \centering
    \includegraphics[width=\columnwidth]{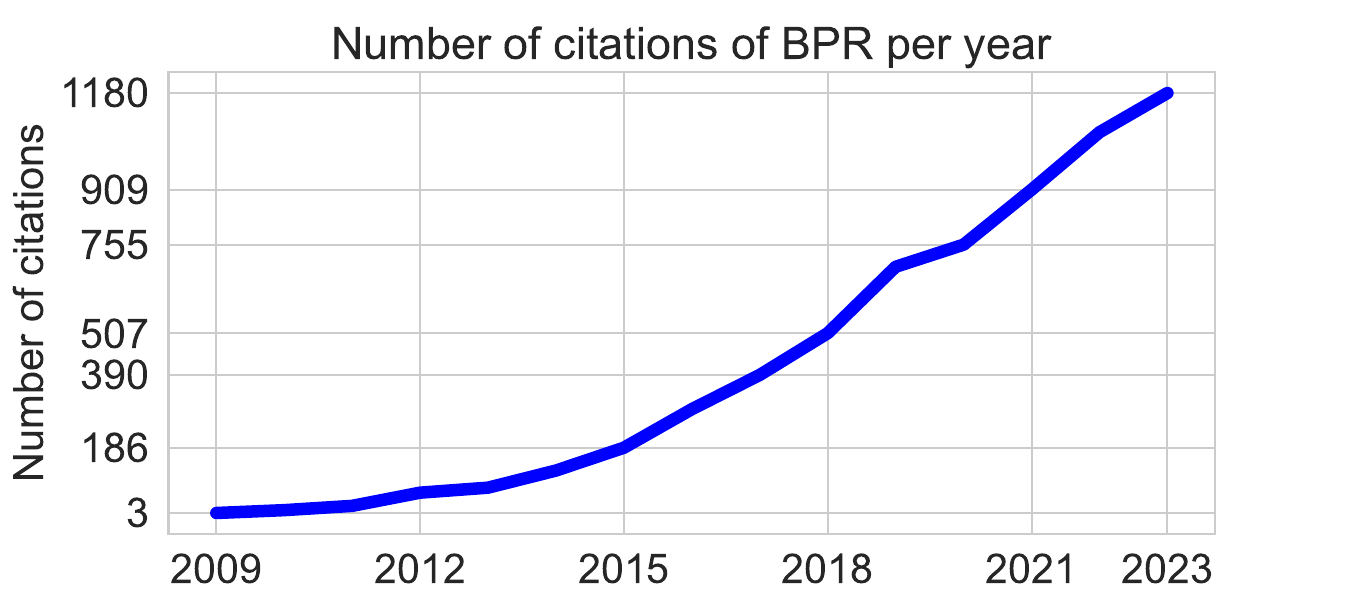}
    \caption{Number of citations of BPR per year according to Google Scholar, as of May 15, 2024.}
    \label{fig:citations}
    \vspace{-1em}
\end{figure}

The advancements in simple linear models and matrix factorization techniques have played a crucial role in the development of recommender systems, a subfield of machine learning.
In recent years, however, deep learning (DL) methods \cite{lecun2015deep} have become dominant in other areas of machine learning, such as natural language processing (NLP) \cite{cambria2014jumping}, computer vision (CV) \cite{voulodimos2018deep}, and reinforcement learning (RL) \cite{kaelbling1996reinforcement}. Given these advances, one might expect similar improvements in recommender systems. Contrarily, several studies on recommender systems suggest that linear models are more effective than DL models for top-n recommendation \cite{shevchenko2024variability, ferrari2019we, li2023next}. These linear methods are easier and faster to train than many deep learning models. Additionally, many neural network models that claim superiority in specific tasks over simple linear models often face reproducibility issues, as outlined in \cite{petrov2022systematic, hidasi2023gru4rec}. 

\looseness -1 An excellent example of a linear model, proven successful in many recommendation systems tasks, is the Bayesian Personalised Ranking (BPR) model \cite{BPR}. The authors introduced pairwise ranking loss in matrix factorization, arguing that it is better suited for item recommendation tasks than pointwise loss functions, such as quadratic regression loss functions \cite{hu2008collaborative}. The idea of a pairwise target in BPR has attracted significant academic attention, evidenced by the increasing number of BPR paper citations, as highlighted in Figure \ref{fig:citations}. With 6624 citations as of May 15, 2024, the BPR paper is one of the most cited in the field of recommender systems.

Two notable factors affect the citation rate of the BPR paper and its popularity. \textit{First}, BPR is a highly extendable model. Many researchers have built upon it with additional ideas, such as GBPR \cite{pan2013gbpr}, VBPR \cite{he2016vbpr}, and CPLR \cite{liu2018cplr}. In contrast, others have modified it to address issues like popularity bias \cite{damak2021debiased} and item fairness \cite{chen2023improving}. \textit{Second}, numerous studies utilize the BPR loss as an objective function for other model architectures. Examples of such modifications include LightGCN \cite{he2020lightgcn},  GRU4Rec \cite{hidasi2018recurrent}, FPMC \cite{rendle2010factorizing}, which incorporate different structures such as graphs, recurrent neural networks, and Markov chains, respectively.

\looseness -1 The widespread popularity of BPR has led to numerous third-party implementations in well-known open-source frameworks such as Implicit \cite{implicit}, Cornac \cite{salah2020cornac}, LightFM \cite{kula2015metadata}, RecBole \cite{recbole[1.2.0]}, and Elliot \cite{elliot}, despite the availability of the trustworthy implementation in MyMediaLite \cite{Gantner2011MyMediaLite} library, provided by the authors of the BPR paper \cite{BPR}\footnote{We find this implementation of BPR to be the most credible one because the original authors are involved in its development.}. Unfortunately, the original implementation is implemented in C\#, which hinders its adoption and integration with Python-based algorithms, which are dominant in the recommender systems field. In addition, it does not support GPU acceleration, making it slower than versions that benefit from it. These factors have contributed to the accumulation of the BPR third-party implementations.

However, recent reproducibility studies \cite{hidasi2023gru4rec, petrov2022systematic} reveal that many third-party implementations for linear and deep learning models lack essential features, complicating comparisons with the original implementations. Thus, despite its widespread adoption, the BPR model might face similar challenges. Moreover, careful tuning of hyperparameters in well-established matrix factorization-based models has been shown to achieve performance near state-of-the-art (SOTA) methods \cite{ferrari2019we, rendle2021revisiting}. This aspect further emphasizes the importance of a thorough evaluation of BPR, as many recent papers report a subpar performance of this model in several evaluation settings.

\looseness -1 Despite recent reviews of several popular models \cite{rendle2021revisiting, hidasi2023gru4rec, petrov2022systematic, ferrari2019we}, a comprehensive reproducibility study of the BPR model has yet to be conducted. The literature employing BPR as a baseline often omits a detailed description of its implementation, with key features such as sampling methods, learnable item biases, optimizer selection, and separate regularization factors missing from many open-source frameworks. These factors could impact BPR's performance, necessitating further investigation into their effects. Furthermore, the current evaluation standard for top-n recommendations relies on a global timeline method with ranking quality metrics \cite{dataleakagestudy}, diverging from the approach outlined in the BPR paper \cite{BPR}. These observations emphasize the demand for a reassessment of BPR to gain deeper insights into its performance and the factors shaping it.

\looseness -1 To address this, we present an extensive set of experiments examining various features of the BPR model, open-source implementations, and datasets. \textit{Our contributions are threefold}. \textbf{First}, we thoroughly investigate open-source implementations of BPR to assess their consistency with the original implementation, revealing differences in the sets of features included. \textbf{Second}, we analyze the importance of each component of BPR on two real-world datasets, offering insights into the model's behavior and performance. \textbf{Third}, through careful hyperparameters tuning, we show that BPR can achieve better results than anticipated by existing implementations, even surpassing SOTA methods in some cases. Particularly, on the Million Song Dataset, the BPR model statistically significantly outperforms Mult-VAE by 10\% in NDCG@100. The source code of the experiments can be found on our GitHub repository\footnote{\url{https://github.com/nemexur/revisit-bpr}}.

\section{Background}

\looseness -1 \noindent \textbf{Top-n Recommendations}. \, Collaborative filtering is a common approach for generating top-n recommendations based on implicit or explicit feedback data \cite{su2009survey}. This method often relies on user-item interaction data and employs various learning methodologies, including rule-based approaches \cite{lin2000collaborative}, heuristics \cite{ji2020re}, and neighborhood methods \cite{wang2006unifying, lops2011content}. The Netflix Prize competition \cite{bennett2007netflix} notably facilitated the development of matrix factorization techniques for recommender systems based on user behavior, leading to the introduction of new methods such as Implicit ALS (iALS) \cite{hu2008collaborative}, SVD++ \cite{svd-plus-plus}.

\looseness -1 Since then, the task of top-n recommendation has seen widespread applications in implicit feedback scenarios. Novel methods continue to emerge, leveraging diverse methods such as linear methods \cite{ning2011slim, steck2019embarrassingly}, variational autoencoders \cite{liang2018variational}, graph convolutional networks \cite{he2020lightgcn}, diffusion models \cite{wang2023diffusion}. Many studies associated with these approaches have demonstrated improvements over previous methods. Notably, the Bayesian Personalized Ranking (BPR) \cite{BPR}, often employed as a matrix factorization-based baseline, is frequently reported to be surpassed by these models \cite{he2020lightgcn, wang2023diffusion, liang2018variational}.

\looseness -1 Interestingly, recent studies have reported that simple linear models can outperform complex architectures. For instance, one study \cite{ferrari2019we} evaluated several deep learning methods against well-tuned baselines, finding that many linear models, similar to BPR, serve as competitive baselines to deep learning models. Another recent study found that the linear model EASE outperformed numerous state-of-the-art methods across several datasets \cite{shevchenko2024variability, anelli2022top}. Thus, linear methods continue demonstrating robust performance on tasks with implicit feedback despite the latest advancements in deep learning.

\looseness -1 \noindent \textbf{Reproducibility Studies}. \, The study of Dacrema et al. \cite{ferrari2019we} has led to a widespread discussion of reproducibility, creating a new research track that focuses on revisiting previously known methods. This area has gained considerable recognition, with numerous papers submitted each year. For example, the author of the popular sequential model GRU4Rec reviewed various open-source implementations, revealing that third-party implementations often lack certain features and contain bugs, impacting performance \cite{hidasi2023gru4rec}. Another comprehensive study \cite{petrov2022systematic} revisited BERT4Rec \cite{bert4rec}, where authors found that various implementations yielded different quality metrics. Similarly, a study on iALS \cite{hu2008collaborative} reproducibility reported that minor changes to the original implementation greatly improved quality, emphasizing the importance of hyperparameters tuning \cite{rendle2021revisiting}. However, to our knowledge, studies have yet to re-examine BPR despite being a highly cited model and often used as a baseline.

\looseness -1 The development of evaluation protocols has also highlighted issues with data leakage \cite{dataleakagestudy, meng2020exploring, filipovic2020modeling}. Recent studies suggest that global temporal split is one of the most reliable ways of conducting offline experiments \cite{meng2020exploring}, and an evaluation protocol without it might lead to incorrect conclusions \cite{dataleakagestudy}. Consequently, assessing BPR in a global temporal split evaluation setup is necessary against the most advanced top-n recommendation methods.

\section{Bayesian Personalized Ranking}

\subsection{Matrix Factorization}
Matrix Factorization (MF) models operate by factoring the user-item rating matrix $\mathit{R}: \mathit{U} \times \mathit{I}$, where $\mathit{U}$ is the set of all users and $\mathit{I}$ is the set of all items. They associate each user and item with distinct sets of features, represented as real-valued vectors or latent factors. Each latent factor, denoted as $p_u$ for users and $q_i$ for items, comprises an $f$-dimensional vector $\mathbb{R}^f$. To estimate the interaction $r_{ui}$ within the matrix $\mathit{R}$, MF computes an inner product $\hat{r}(u, i) = p_uq_i^T \label{eq:r_ui}$   between the corresponding latent factors.
\begin{gather}
    \argmin_{P, Q} \sum_{(u,i) \in \mathit{R}} L(\hat{r}(u, i), r_{ui}) \label{eq:mf-objective}
\end{gather}
The model parameters of MF are the embedding matrices P and Q, which are learned by minimizing the objective function $L$ (\ref{eq:mf-objective}).

Objective functions to learn MF model parameters can be categorized into three groups: pointwise, pairwise, and listwise \cite{learningtorank}.

\subsubsection{Pointwise} The pointwise approach estimates the discrepancy between the predicted score $\hat{r}(u, i)$ and the actual score $r_{ui}$. It may be represented in the form of a squared loss objective:
\begin{equation}
    \argmin_{P, Q} \sum_{(u,i) \in \mathit{R}} (r_{ui} - \hat{r}(u, i))^2 + \Omega(P, Q)
\end{equation}
\looseness -1 Here, the second term, $\Omega$, penalizes the complexity of the model, often represented as an L2 regularization. The main limitation of pointwise approaches is the demand for a direct item relevance estimation instead of the item comparison for a user, which is a crucial aspect in recommendation tasks.

\looseness -1 \subsubsection{Pairwise} The pairwise objective function aims to optimize the relative ranking of items for a user, in contrast to predicting individual interactions in pointwise approaches, which addresses one of the main limitations of the previous approach. The main idea behind it is to correctly order pairs of items within the context of a user, achieved by comparing items in the user history $\mathit{I}^+(u) = \{i : (u, i) \in \mathit{R}\}$ with the remaining items $j \in I \setminus I^+(u)$. The objective function (\ref{eq:pairwise-obj}) then penalizes the model if the item $j$ is ranked higher than the selected item from the user history using the function $\phi$.
\begin{gather}
    (u, i, j) \in \mathit{D_R}: \Leftrightarrow i \in \mathit{I}^+(u) \wedge j \in I \setminus I^+(u)\\
    \argmin_{P, Q} \sum_{(u,i,j) \in \mathit{D_R}} \phi(\hat{r}(u, i) - \hat{r}(u, j)) + \Omega(P, Q) \label{eq:pairwise-obj}
\end{gather}

\subsubsection{Listwise} 
\looseness -1  The listwise group of functions looks at the whole ranked list of items for each user $\mathbf{r}_u = [r_{u1}, \dots, r_{uI}]$ and optimizes it based on the ranking metric. One significant advantage of the listwise approach over other methods is its ability to distinguish the position of an item within the ranked list. It aids the model in understanding that items at the top of the list are more important than those at the bottom, which is lacking in the pairwise approach. However, these objective functions (\ref{eq:listwise-obj}) are more computationally extensive due to the direct optimization of the ranking quality of the entire recommendation list for a user $\mathbf{\hat{r}}_u = [\hat{r}(u, 1), \dots, \hat{r}(u, I)]$.
\begin{gather}
    \argmin_{P, Q} \sum_{u \in \mathit{U}} L(\mathbf{\hat{r}}_u, \mathbf{r}_u) + \Omega(P, Q) \label{eq:listwise-obj}
\end{gather}

\subsection{Bayesian Personalized Ranking Features}

Bayesian Personalized Ranking (BPR) \cite{BPR} is a popular version of the matrix factorization model with a pairwise objective function. What sets BPR apart from other pairwise models is its objective function. The goal of the function is to minimize the log-likelihood of the correct ordering in the form of:

\begin{equation}
    \argmin_{P, Q} \sum_{(u,i,j) \in \mathit{D_R}} - \log\sigma(\hat{r}(u, i) - \hat{r}(u, j))
    \label{eq:bpr-obj}
\end{equation}

\looseness -1 During the training phase, it receives the tuple of a user $u$, a positive item $i$, and a negative item $j$ and then computes the objective (\ref{eq:bpr-obj}) using the function $\hat{r}(.)$. While $\hat{r}(.)$ frequently utilizes a Matrix Factorization model, it can be represented by any algorithm that calculates the score between a user and an item, such as KNN algorithms and Weighted MF utilized in \cite{BPR}. Although the choice of the scoring function impacts the model's performance, we adhere to the matrix factorization version for our experiments. Additionally, five more features of BPR might influence its behavior:

\noindent \textbf{Regularization}. \, Authors of the original BPR paper \cite{BPR} introduced three separate regularization parameters: user regularization, positive item regularization, and negative item regularization in the form of $\Omega(P, Q) = \sum_{(u,i,j) \in \mathit{D_R}} \lambda_u p_u^2 + \lambda_i q_i^2 + \lambda_j q_j^2$, where $\lambda_.$ are regularization terms for specific embeddings.

\noindent \textbf{Optimizer}. \, The original implementation utilizes standard Stochastic Gradient Descent (SGD) for training. Hence, it is reasonable to expect potential performance improvements by integrating the latest optimization techniques, such as Momentum SGD \cite{sgd_momentum}, RMSProp \cite{RMSProp}, and Adam \cite{kingma2017adam}.

\noindent \textbf{Negative Sampling}. \, In \cite{bprv2}, the authors of BPR observed that uniform sampling is not the best choice of negative sampling algorithm for the model. As the training progresses, it starts to yield \textit{easy} negatives that result in minimal model parameter changes. To address this issue, they introduced the \textit{adaptive sampling} algorithm for negatives, which respects the model's performance at each iteration. Although this sampling algorithm is not a part of the original paper \cite{BPR}, we still consider it a necessary addition to the algorithm that should be included in the list.

\looseness -1 \noindent \textbf{Item Bias}. \, Many popular model architectures introduce item biases to the model as a prominent feature that frequently improves the performance \cite{bert4rec, cdae}. In the case of MF, the addition of item biases changes the inner product to $\hat{r}(u, i) = b_i + p_uq_i^T \label{eq:r_ui-biased}$. Furthermore, as we will show in Section \ref{section:bpr-impls}, these biases exist in many popular open-source BPR implementations. Notably, the original implementation in MyMediaLite\footnote{\url{https://github.com/zenogantner/MyMediaLite}} \cite{Gantner2011MyMediaLite} also includes learnable item biases.

\looseness -1 \noindent \textbf{User Bias}. \, User biases $\hat{r}(u, i) = b_u + p_uq_i^T$, like item biases, might also impact the model's performance. However, we have excluded them from the experiments because they are uncommon in open-source BPR implementations.

\section{Experiments}

We conduct the experiments to address the following research questions:
\begin{enumerate}[font={\bfseries}, label={RQ\arabic*}]
    \item How do the results achieved using open-source implementations compare with the originally reported results?
    \item How do the BPR model's features help improve performance?
    \item How does the fine-tuned BPR compare to state-of-the-art models in the top-n recommendations?
\end{enumerate}

\subsection{Experimental Setup}

\begin{table}[t]
\caption{Statistics for the complete training datasets after preprocessing. The median per user/item is the middle of the distribution of the number of actions for users or items.}
\label{table:datasets-stats}
\resizebox{\linewidth}{!}{
\begin{tabular}{lrrrrrr}
\toprule
Dataset                                                       & Users  & Items & Actions & Sparsity & Med. User/Item \\ \midrule
Netflix\footnotemark                                          & 9949   & 4825  & 563577  & 0,9883   & 27/12          \\
ML-20M                                                        & 136677 & 20108 & 9,7M    & 0,9965   & 37/16          \\
MSD                                                           & 571355 & 41140 & 32,5M   & 0,9986   & 39/383         \\
Yelp (time-split)                                             & 252616 & 92089 & 2,2M    & 0,9999   & 5/8            \\
\begin{tabular}[c]{@{}l@{}}ML-20M\\(time-split)\end{tabular}  & 124377 & 12936 & 8.9M    & 0,9944   & 38/57          \\
\bottomrule
\end{tabular}
}
\footnotesize{$^4$the subsample of the dataset similar to the original paper \cite{BPR}.}
\end{table}

\looseness -1 \subsubsection{Datasets} Similar to \cite{rendle2021revisiting}, we experiment using the benchmarks established in \cite{liang2018variational}. Furthermore, we include evaluation protocols with a global time-based split proposed in \cite{hidasi2023evalflaws}. The experiments are conducted over four publicly available datasets: MovieLens-20M (\textbf{ML-20M}) \cite{movielens}, Million Song Dataset (\textbf{MSD}) \cite{msd}, Netflix \cite{bennett2007netflix}, and Yelp \cite{asghar2016yelp}. The characteristics of the datasets are summarized in Table \ref{table:datasets-stats}.

\subsubsection{Preprocessing and Evaluation Protocol} The evaluation methodology adopted in this study utilizes three methods, depending on the described research question.

\looseness -1 \noindent \textbf{RQ1}. \, We employ the preprocessing and evaluation protocol from the BPR paper \cite{BPR} on the Netflix dataset for this RQ. In this paper, the authors subsampled the dataset after preprocessing. We follow the same strategy and sample the dataset with a similar number of interactions: 563000 samples compared to 565000 samples in \cite{BPR}.

\looseness -1 \noindent \textbf{RQ2}. \, We opt for ML-20M and MSD to conduct experiments for this RQ. We follow the same preprocessing and evaluation procedures from \cite{steck2019embarrassingly, liang2018variational}. For detailed steps and description, refer to \cite{liang2018variational} and their code\footnote{\url{https://github.com/dawenl/vae_cf}}. However, we slightly modify it by adding held-out users to the training part, as the BPR model cannot handle cold users. The validation and testing parts remain unchanged.

\looseness -1 \noindent \textbf{RQ3}. \, In addition to the datasets used in RQ2, we conduct experiments using a global temporal split on ML-20M and Yelp datasets, following \cite{hidasi2023evalflaws, dataleakagestudy}. To preprocess these datasets, we first convert all numeric ratings or the presence of an interaction into implicit feedback. Afterward, we keep users and items with at least three interactions to ensure the dataset's quality, following \cite{he2017neural, kang2018selfattentive, rendle2010factorizing}. Regarding evaluation protocol, we set the testing time window to three years and the validation time window to one year for both datasets. Additionally, we filter out users absent from training and complete training parts in validation and testing on each dataset, respectively.

\subsubsection{Metrics} We employ two ranking-based metrics: \textit{Recall@K} (R) and \textit{NDCG@K} (N) with binary relevance function. Following \cite{rendle2021revisiting}, for each user, we remove all items from the predictions that the user has already interacted with in the training dataset.

\begin{table}[t]
\caption{BPR implementations. Source of the implementation: from the ACM Recommender Systems List of Frameworks ($\star$), from the GitHub search ($\diamondsuit$). Availability of features: available (\cmark), missing (\xmark). Original - MyMediaLite implementation.}
\label{table:bpr-impls}
\resizebox{\linewidth}{!}{
\begin{tabular}{l|ll|c|cccc|cc|cc}
\toprule
\multirow{7}{*}{}                                                                 & \multirow{7}{*}{\R{GitHub Stars\footnotemark}} & \multirow{7}{*}{\R{Commits\footnotemark}} & \multicolumn{9}{c}{\textbf{Features}}                                                                                                                                                                                                                                                                                                                           \\ \cline{4-12}
                                                                                  &                                                &                                           & \multicolumn{1}{c|}{\multirow{5}{*}{\begin{tabular}[c]{@{}c@{}}Item\\ Biases\end{tabular}}} & \multicolumn{4}{c|}{Regularization}                                                                                    & \multicolumn{2}{c|}{Optimizer}                          & \multicolumn{2}{c}{\begin{tabular}[c]{@{}c@{}}Negative\\Sampling\end{tabular}} \\ \cline{5-12}
                                                                                  &                                                &                                           & \multicolumn{1}{c|}{}                                                                       & \R{User $\lambda_u$} & \R{Positive $\lambda_i$} & \R{Negative $\lambda_j$} & \multicolumn{1}{c|}{\R{Shared $\lambda$}} & \R{SGD}  & \multicolumn{1}{c|}{\R{Others\footnotemark}} & \R{Uniform}         & \R{Adaptive}                                             \\
\midrule                                                                                                                           
\href{https://github.com/PreferredAI/cornac}{$\text{Cornac}^\star$}               & 820                                            & 116 \C{86 2023, 30 2024}                  & \cmark                                                                                      & \xmark               & \xmark                   & \xmark                   & \cmark                                    & \cmark   & \xmark                                       & \cmark              & \xmark                                                   \\
\href{https://github.com/recsys-benchmark/DaisyRec-v2.0}{$\text{DaisyRec}^\star$} & 55                                             & 31 \C{31 2023, 0 2024}                    & \xmark                                                                                      & \xmark               & \xmark                   & \xmark                   & \cmark                                    & \cmark   & \cmark                                       & \cmark              & \xmark                                                   \\
\href{https://github.com/sisinflab/elliot}{$\text{Elliot}^\star$}                 & 265                                            & 1 \C{1 2023, 0 2024}                      & \cmark                                                                                      & \cmark               & \cmark                   & \cmark                   & \xmark                                    & \cmark   & \xmark                                       & \cmark              & \xmark                                                   \\
\href{https://github.com/RUCAIBox/RecBole}{$\text{Recbole}^\star$}                & 3,200                                          & 174 \C{164 2023, 10 2024}                 & \xmark                                                                                      & \xmark               & \xmark                   & \xmark                   & \cmark                                    & \cmark   & \cmark                                       & \cmark              & \xmark                                                   \\
\href{https://github.com/THUwangcy/ReChorus}{$\text{ReChorus}^\star$}             & 492                                            & 5 \C{5 2023, 0 2024}                      & \xmark                                                                                      & \xmark               & \xmark                   & \xmark                   & \cmark                                    & \cmark   & \cmark                                       & \cmark              & \xmark                                                   \\
\href{https://gitlab.com/recpack-maintainers/recpack}{$\text{RecPack}^\star$}     & 6                                              & 70 \C{70 2023, 0 2024}                    & \xmark                                                                                      & \cmark               & \cmark                   & \xmark                   & \xmark                                    & \xmark   & \cmark                                       & \cmark              & \xmark                                                   \\
\href{https://github.com/benfred/implicit}{$\text{Implicit}^\diamondsuit$}        & 3,400                                          & 27 \C{27 2023, 0 2024}                    & \cmark                                                                                      & \xmark               & \xmark                   & \xmark                   & \cmark                                    & \cmark   & \xmark                                       & \xmark\footnotemark & \xmark                                                   \\
\href{https://github.com/lyst/lightfm}{$\text{LightFM}^\diamondsuit$}             & 4,600                                          & 15 \C{15 2024, 0 2024}                    & \cmark                                                                                      & \cmark               & \cmark                   & \xmark                   & \xmark                                    & \xmark   & \cmark                                       & \cmark              & \xmark                                                   \\
\href{https://github.com/zenogantner/MyMediaLite}{$\text{\textbf{Original}}$}     & 498                                            & 0 \C{0 2023, 0 2024}                      & \cmark                                                                                      & \cmark               & \cmark                   & \cmark                   & \xmark                                    & \cmark   & \xmark                                       & \cmark              & \xmark                                                   \\ \hline
\multicolumn{3}{c|}{\textbf{Ours}}                                                                                                                                             & \cmark                                                                                      & \cmark               & \cmark                   & \cmark                   & \cmark                                    & \cmark   & \cmark                                       & \cmark              & \cmark                                                   \\ 
\bottomrule
\end{tabular}
}
\footnotesize{$^6$As of May 15, 2024. $^7$From January 1, 2023, to May 15, 2024. $^8$Adam, Adagrad, Momentum SGD, RMSProp. $^9$Implicit utilizes popularity-based negative sampling.}
\end{table}

\looseness -1 \subsubsection{BPR Implementations} \label{section:bpr-impls} To identify open-source implementations of BPR, we looked through the list of recommender frameworks by ACM Recommender Systems Conference\footnote{\url{https://github.com/ACMRecSys/recsys-evaluation-frameworks}} and the list of popular recommender systems libraries on GitHub not present in that list. In Table \ref{table:bpr-impls}, one can find frameworks with the BPR model and which features they implement. Notably, Microsoft Recommenders\footnote{\url{https://github.com/recommenders-team/recommenders}} is absent from this list because the authors of this repository use Cornac \cite{salah2020cornac} to implement the BPR model.

\looseness -1 Our selection criteria for experiments prioritize open-source implementations with good quality and support, assessed through the number of GitHub stars and consistent maintenance efforts. Specifically, we combine the Top-5 implementations based on GitHub stars with the Top-5 based on the number of commits. Additionally, we include MyMediaLite, regarded as the original implementation of BPR used for experiments in \cite{BPR}, and Elliot, as it closely follows the original BPR model from \cite{BPR}, as observed in Table \ref{table:bpr-impls}. Based on these criteria, we select the following implementations: Cornac, Elliot, Implicit, LightFM, Recbole, and MyMediaLite (\textbf{Original}).

\looseness -1 Additionally, it is evident from Table \ref{table:bpr-impls} that only Elliot \cite{elliot} implements the complete set of features of the original model. Most other implementations use shared regularization factors rather than separate ones for users, positive and negative items. Moreover, all implementations, including the original, lack adaptive negative sampling from \cite{bprv2}, which is a crucial feature, as we will demonstrate later in Section \ref{section:rq2}. Interestingly, many implementations introduce item biases as an additional feature, diverging from the original paper \cite{BPR}.

\looseness -1 Moreover, as we will show in Section \ref{section:rq1}, Cornac performs best among third-party BPR implementations. However, it has a notable limitation: Cornac utilizes Cython, which only supports CPU execution, resulting in longer training time than GPU alternatives. Moreover, we could not add adaptive negative sampling to Cornac's BPR model due to Cython constraints with sampling algorithms. Therefore, we decided to replicate BPR with PyTorch \cite{paszke2019pytorch}, which supports GPU, allowing us to incorporate all discussed features completely.

\subsubsection{Baselines} In addition, we compare BPR implementations to baselines that have demonstrated state-of-the-art results in recent reproducibility papers on the selected datasets:
\begin{itemize}
    \item \textbf{ItemPop}. A non-personalized model that ranks items based on their popularity. This baseline is used for reference.
    \item \textbf{EASE} \cite{steck2019embarrassingly}. An item-item collaborative filtering model with a closed-form solution. A state-of-the-art model on MSD \cite{shevchenko2024variability}.
    \item \textbf{Mult-VAE} \cite{liang2018variational}. An extension of variational autoencoders for collaborative filtering with implicit feedback using a multinomial likelihood objective. In addition, we include \textbf{Mult-DAE} as this version might outperform Mult-VAE in the most active users, which comprises the biggest part of any dataset.
\end{itemize}

\subsubsection{Hyperparameters Search} Hyperparameters are optimized on a dedicated training/validation split created from the complete training set using the same process as the train/test split. Subsequently, the models are retrained on the complete training dataset with the best hyperparameters and evaluated on the testing dataset. The optimization procedure uses a popular framework for hyperparameters optimization Optuna \cite{akiba2019optuna} on all datasets using the sampler based on the Tree-structured Parzen Estimator (TPE) \cite{tpe} algorithm. We search for the best hyperparameters using NDCG@100 on all datasets except Netflix, which utilizes the AUC metric.

Regarding model-specific hyperparameters, we adjust the L2-norm regularization parameter for the EASE model \cite{steck2019embarrassingly}. In the case of the Mult-VAE and Mult-DAE models, we vary the number of epochs and the learning rate while keeping the batch size consistent with the original paper \cite{liang2018variational}. The considered hyperparameter ranges and distributions are available on our GitHub repository.

\looseness -1 \subsubsection{Embedding Dimension} Matrix factorization models rely on the embedding dimension of user and item representations. It controls the capacity of the model. Models with high capacity are more likely to overfit the training set, whereas models with low capacity may underfit it. To explore the impact of embedding dimensions following \cite{rendle2021revisiting}, we conduct the experiments with varying dimensions: 32, 64, 128, 256, 512, and 1024 for ML-20M and MSD datasets, 16, 32, 64, and 128 for Netflix dataset, and 64, 128, 256, and 512 for ML-20M time-splitted and Yelp. We run the full hyperparameters search with fixed embedding dimension to properly asses the impact on the BPR algorithm. This approach allows us to identify the optimal embedding dimension for each dataset, which we then utilize in subsequent experiments on the dataset.

\begin{figure}[t]
\centering
\includegraphics[width=9cm]{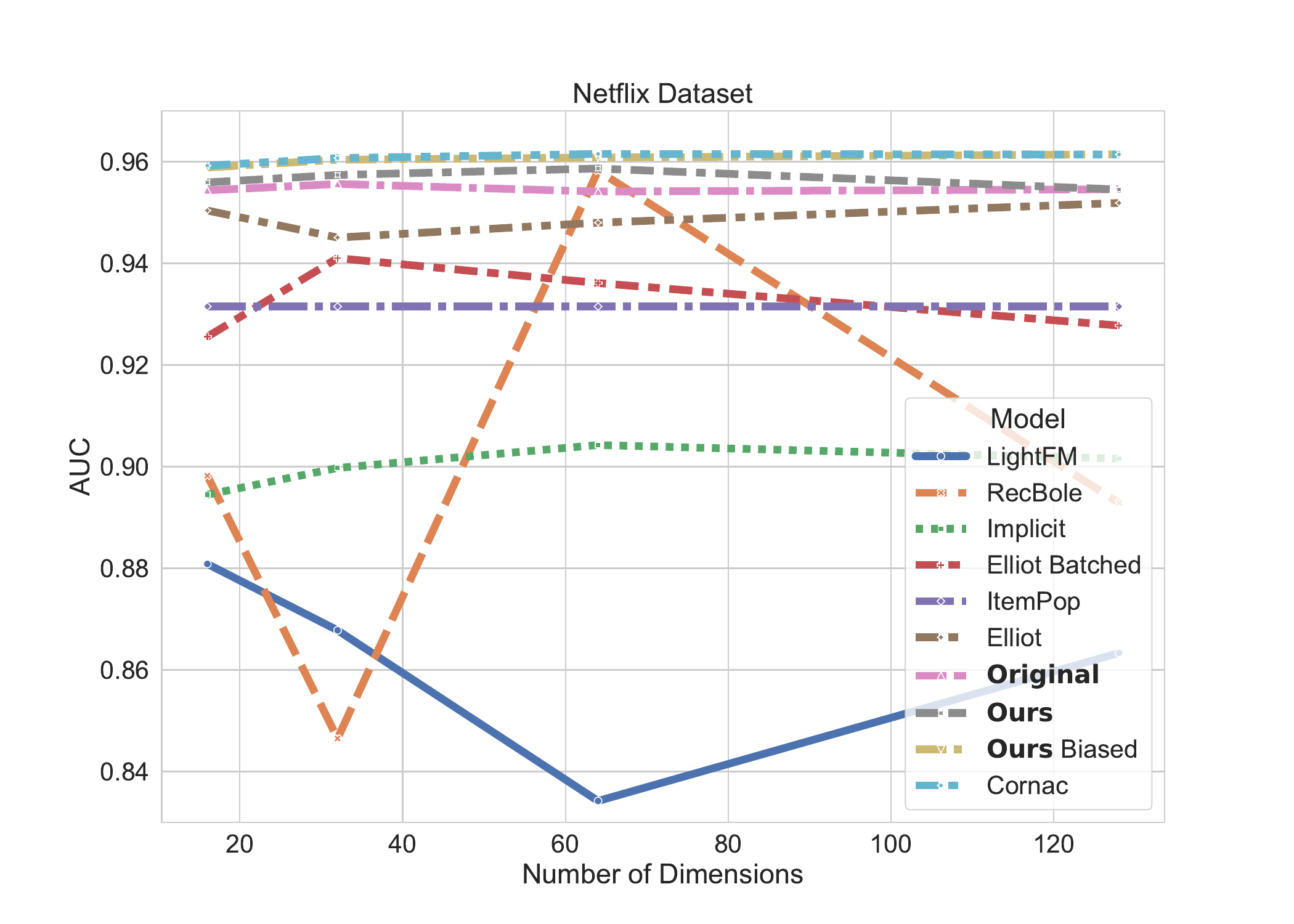}
\caption{Area Under the ROC Curve (AUC) prediction quality for the Netflix dataset using various open-source BPR implementations. The ItemPop model is included to compare performance against a non-personalized baseline.}
\label{fig:orig-bpr-impls}
\end{figure}

\section{Experimental Results}
\subsection{RQ1. Replicability} \label{section:rq1}

First, we analyze whether the selected open-source implementations can replicate the results of MyMediaLite BPR implementation (\textbf{Original}), a trustworthy version of BPR from the original authors, in a setup as close as possible to the original BPR paper. Figure \ref{fig:orig-bpr-impls} compares the implementations with the default configuration and hyperparameters search over epochs, regularization, and learning rate for each implementation. As shown in Figure \ref{fig:orig-bpr-impls}, our implementation, Cornac's, and Elliot's implementations are close to the original results. Conversely, other open-source implementations exhibit inferior results compared to other models. Interestingly, Recbole and LightFM demonstrate unstable performance levels regarding the AUC metric in Figure \ref{fig:orig-bpr-impls}. We assume this behavior is related to the adaptive optimization algorithms used in these implementations. Specifically, Adam \cite{kingma2017adam} in RecBole and Adagrad \cite{adagrad} in LightFM. We observe a similar pattern in the experiments for Section \ref{section:rq2}, where we employ adaptive optimizers. We will provide an explanation of these findings in the subsequent sections.

\looseness -1 Additionally, we check two implementations of our model: one incorporating item biases and one without them. Notably, on the Netflix dataset, we observe that item biases are only helpful at the start of training, but their utility diminishes over time. Eventually, non-biased models reach the same performance levels as biased models, being less than 1\% worse than their counterparts.

\looseness -1 Moreover, we assess two implementations of BPR in Elliot framework: BPRMF\footnote{\url{https://github.com/sisinflab/elliot/blob/v0.3.1/elliot/recommender/latent_factor_models/BPRMF/BPRMF_model.py}} and Batched BPRMF\footnote{\url{https://github.com/sisinflab/elliot/blob/v0.3.1/elliot/recommender/latent_factor_models/BPRMF_batch/BPRMF_batch_model.py}}. The former closely follows the BPR model from \cite{BPR}, while the latter is an optimized variant that diverges from the original BPR model. Specifically, Batched BPR is similar to LightFM but utilizes SGD instead of Adagrad optimizer. Therefore, we suppose these changes result in a 3\% drop in performance compared to the BPRMF variant from Elliot.

\subsection{RQ2. Influence of BPR features} \label{section:rq2}

\begin{figure*}[t]
\centering
\subcaptionbox{MovieLens-20M --- NDCG@100}{\includegraphics[width=8cm]{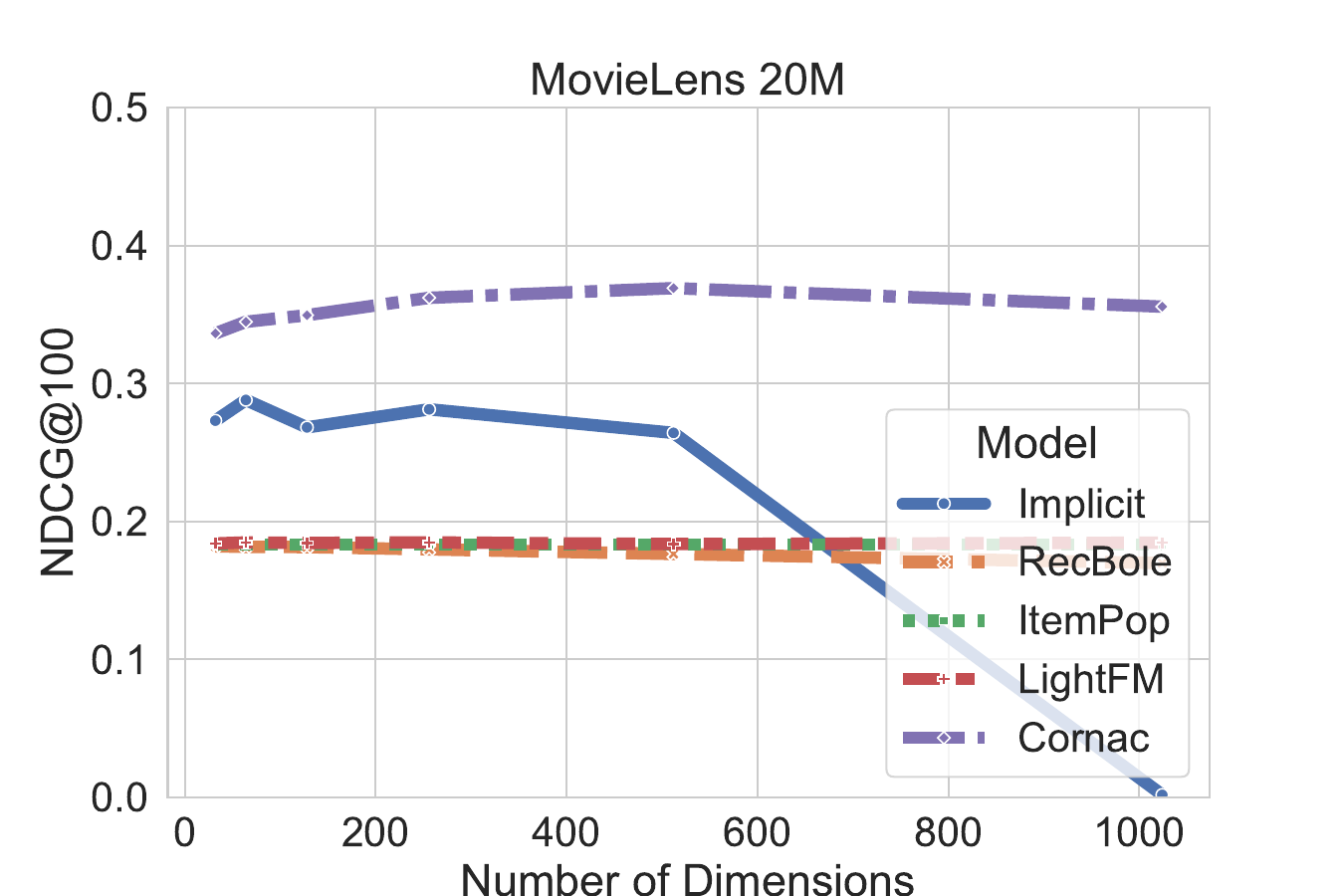}}%
\subcaptionbox{MSD --- NDCG@100}{\includegraphics[width=8cm]{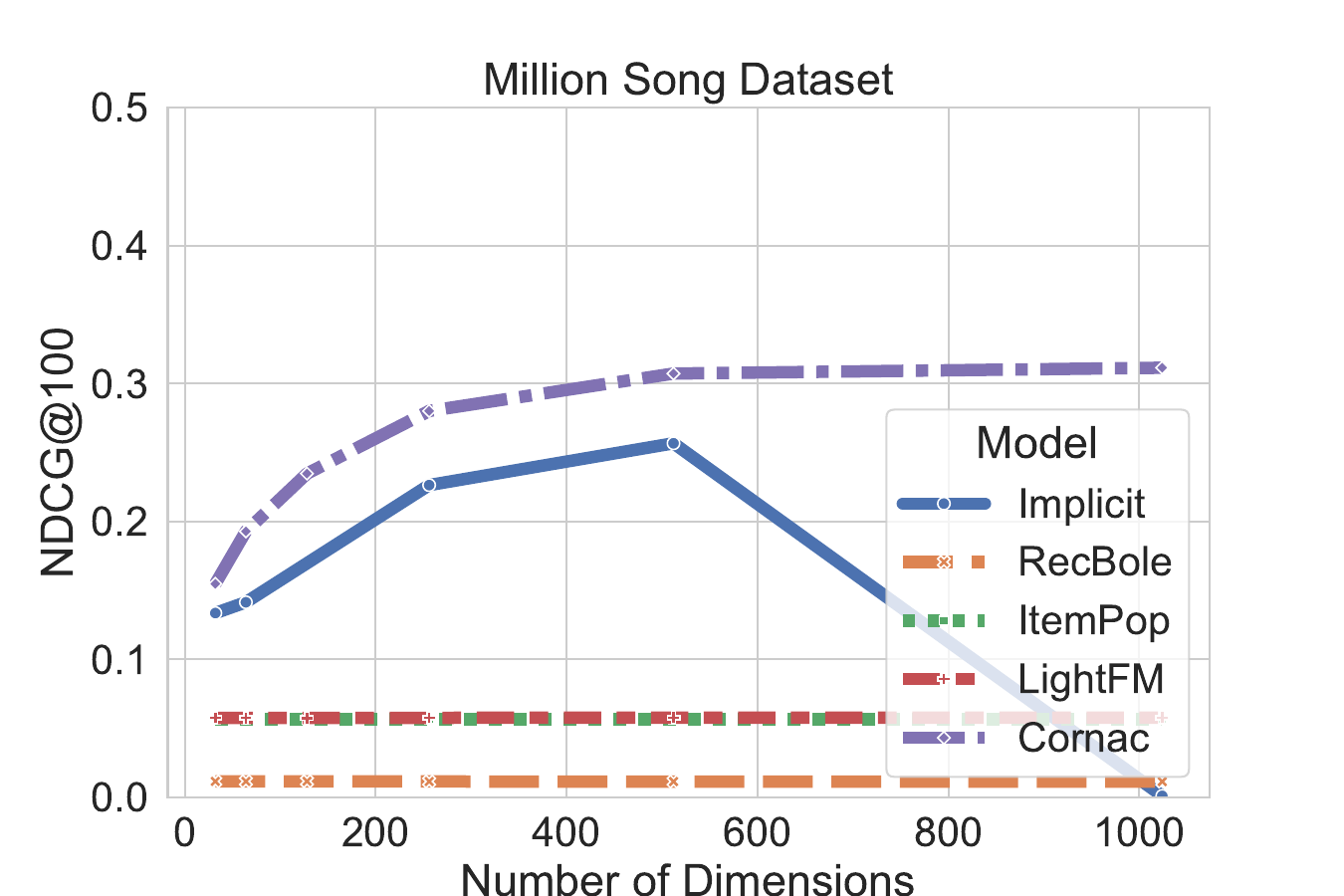}}%
\caption{Performance in NDCG@100 relative to the number of embedding dimensions on the two datasets.}
\label{fig:bpr-impls-emb-dim}
\end{figure*}

\begin{table}[t]
\caption{The list of values for each subset of BPR features we use in the experiments. Bold denotes the best set of features according to our experiments in Section \ref{section:rq2} for ML-20M.}
\label{table:bpr-features-grid}
\begin{tabular}{c|c}
\toprule
\textbf{Feature}                   & \textbf{Values}                                                                                  \\ \midrule   
\multirow{2}{*}{Item Biases}       & Enabled                                                                                          \\
                                   & \textbf{Disabled}                                                                                \\ \hline
\multirow{4}{*}{Regularization}    & \boldmath$\lambda_u p_u^2 + \lambda_i q_i^2 + \lambda_j q_j^2 (\lambda_u, \lambda_i, \lambda_j)$ \\
                                   & $\lambda_u p_u^2 + \lambda_i q_i^2 + \lambda_i q_j^2 (\lambda_u, \lambda_i)$                     \\
                                   & $\lambda p_u^2 + \lambda q_i^2 + \lambda q_j^2 (\lambda)$                                        \\
                                   & No Regularization                                                                                \\ \hline
\multirow{4}{*}{Optimizer}         & \textbf{SGD}                                                                                     \\
                                   & Momentum SGD                                                                                     \\
                                   & RMSprop                                                                                          \\
                                   & Adam                                                                                             \\ \hline
\multirow{2}{*}{Negative Sampling} & Uniform                                                                                          \\ 
                                   & \textbf{Adaptive}                                                                                \\
\bottomrule
\end{tabular}
\end{table}

\looseness -1 In Table \ref{table:bpr-impls}, we outline the availability of BPR features in each open-source BPR implementation. In order to demonstrate the effect of these features on model's performance, we conduct experiments comparing various subsets of the features using our implementation: (1) item biases, (2) regularization approaches, (3) optimizers, and (4) negative sampling. To speed up the hyperparameters search, we first fit our models with fewer epochs and then train the best model with the best hyperparameters using more epochs. Specifically, we opt for 70 epochs for the hyperparameters search phase, 1000 epochs for the best model on the ML-20M dataset, and 200 epochs for the MSD dataset. Such a two-step approach for hyperparameters search proves efficient for the BPR model, as it continues to train after thousands of epochs with minimal overfitting, mainly when uniform-based sampling produces good negative samples according to \cite{bprv2}. Additionally, the training protocol employs the Early Stopping criterion with patience equal to 13 epochs.

We conduct experiments following the grid of possible values for each feature outlined in Table \ref{table:bpr-features-grid}. The most comprehensive method for experimenting with these features involves individually testing each combination of values. However, the cardinality of this approach is 64 experiments on each dataset, which results in over 5000 models. In order to lower the number of combinations, we split the features into three subsets: (1) item biases and regularization, (2) optimizers, and (3) negative sampling. Notably, \textit{the first subset} holds the highest importance, as the optimal combination of values identified here serves as the base model for subsequent subsets. It lowers the number of experiments to 14 for each dataset (over 1000 models) but adds the sequential order to them.

Furthermore, we plan to evaluate the impact of the embedding dimension on the model in isolation. This modification significantly increases the number of trained models, particularly by a factor of six, thereby considerably prolonging the experiment's duration. To streamline this, we obtain a rough estimate of the embedding dimension's influence from open-source implementations. We conduct a full hyperparameters search for the Cornac, Implicit, LightFM, and RecBole libraries, using all available hyperparameters for the BPR model provided by these implementations. We excluded MyMediaLite and Elliot from the list due to their significantly longer training times than the other open-source implementations.

\looseness -1 Figure \ref{fig:bpr-impls-emb-dim} summarizes the results of this estimate. It is evident that Cornac achieves the best results with embedding dimensions of 512 and 1024 on both datasets. Surprisingly, there is a huge performance drop when Implicit employs embedding dimensions of size 1024. We assume this behavior is related to popularity-based negative sampling utilized in BPR implementation for Implicit, which might hinder the model's performance, as highlighted in \cite{bprv2}. LightFM and Recbole implementations exhibit even lower performance than Implicit. We observed that both LightFM and Recbole adapt uniform negative sampling, but there is a discrepancy in their optimization algorithms. Cornac and Implicit implementations adapt regular Stochastic Gradient Descent (SGD), whereas LightFM and Recbole employ Adagrad \cite{adagrad} and Adam \cite{kingma2017adam}, respectively. We suppose that the choice of optimization algorithms led to these results, supporting our initial idea that the optimizer is a vital feature of the BPR model and the findings described in Section \ref{section:rq1}.

\begin{table}[t]
\caption{Evaluation of the BPR model in terms of NDCG@K (N@K) and Recall@K (R@K) using various subsets of item biases and regularization features. All models in the table were trained with the SGD optimizer and uniform negative sampling. The best value is bolded, and the second-best is underlined.}
\label{table:bpr-optimizers}
\resizebox{\linewidth}{!}{
\begin{tabular}{c|cccc|cccccc}
\toprule
\multicolumn{5}{c|}{\textbf{Features}}                                                                                                                                                                               & \multirow{6}{*}{N@5} & \multirow{6}{*}{N@10} & \multirow{6}{*}{N@100} & \multirow{6}{*}{R@5} & \multirow{6}{*}{R@10} & \multirow{6}{*}{R@100} \\ \cline{1-5}
\multicolumn{1}{c|}{\multirow{5}{*}{\begin{tabular}[c]{@{}c@{}}Item\\ Biases\end{tabular}}} & \multicolumn{4}{c|}{Regularization}                                                                                    & \multicolumn{1}{c}{} & \multicolumn{1}{c}{}  &                        &                      &                       &                        \\ \cline{2-5}
\multicolumn{1}{c|}{}                                                                       & \R{User $\lambda_u$} & \R{Positive $\lambda_i$} & \R{Negative $\lambda_j$} & \multicolumn{1}{l|}{\R{Shared $\lambda$}} & \multicolumn{1}{c}{} & \multicolumn{1}{c}{}  &                        &                      &                       &                        \\ \midrule
\multicolumn{11}{c}{ML-20M}                                                                                                                                                                                                                                                                                                                                          \\ \midrule
\xmark                                                                                      & \xmark               & \xmark                   & \xmark                   & \xmark                                    & 0.2537               & 0.2504                & 0.3510                 & 0.1246               & 0.1948                & 0.5735                 \\
\xmark                                                                                      & \xmark               & \xmark                   & \xmark                   & \cmark                                    & 0.2656               & 0.2627                & 0.3665                 & 0.1316               & 0.2048                & 0.5938                 \\
\xmark                                                                                      & \cmark               & \cmark                   & \xmark                   & \xmark                                    & 0.2721               & \underline{0.2695}    & 0.3694                 & \underline{0.1353}   & \underline{0.2100}    & 0.5940                 \\
\xmark                                                                                      & \cmark               & \cmark                   & \cmark                   & \xmark                                    & \textbf{0.2929}      & \textbf{0.2879}       & \textbf{0.3883}        & \textbf{0.1398}      & \textbf{0.2171}       & \textbf{0.6150}        \\
\cmark                                                                                      & \xmark               & \xmark                   & \xmark                   & \xmark                                    & 0.2517               & 0.2469                & 0.3499                 & 0.1185               & 0.1863                & 0.5771                 \\
\cmark                                                                                      & \xmark               & \xmark                   & \xmark                   & \cmark                                    & 0.2419               & 0.2381                & 0.3441                 & 0.1180               & 0.1855                & 0.5707                 \\
\cmark                                                                                      & \cmark               & \cmark                   & \xmark                   & \xmark                                    & 0.2565               & 0.2514                & 0.3530                 & 0.1221               & 0.1906                & 0.5766                 \\
\cmark                                                                                      & \cmark               & \cmark                   & \cmark                   & \xmark                                    & \underline{0.2748}   & 0.2686                & \underline{0.3723}     & 0.1291               & 0.2020                & \underline{0.5996}     \\ \midrule
\multicolumn{11}{c}{MSD}                                                                                                                                                                                                                                                                                                                                             \\ \midrule
\xmark                                                                                      & \xmark               & \xmark                   & \xmark                   & \xmark                                    & 0.1938               & 0.1906                & 0.2753                 & 0.1007               & 0.1543                & 0.4145                 \\
\xmark                                                                                      & \xmark               & \xmark                   & \xmark                   & \cmark                                    & \underline{0.2183}   & \underline{0.2150}    & \underline{0.3115}     & \underline{0.1098}   & \underline{0.1707}    & \textbf{0.4695}        \\
\xmark                                                                                      & \cmark               & \cmark                   & \xmark                   & \xmark                                    & 0.2085               & 0.2057                & 0.3018                 & 0.1052               & 0.1638                & 0.4593                 \\
\xmark                                                                                      & \cmark               & \cmark                   & \cmark                   & \xmark                                    & 0.1844               & 0.1828                & 0.2796                 & 0.0939               & 0.1472                & 0.4396                 \\
\cmark                                                                                      & \xmark               & \xmark                   & \xmark                   & \xmark                                    & 0.1859               & 0.1828                & 0.2706                 & 0.0960               & 0.1477                & 0.4142                 \\
\cmark                                                                                      & \xmark               & \xmark                   & \xmark                   & \cmark                                    & \textbf{0.2249}      & \textbf{0.2196}       & \textbf{0.3132}        & \textbf{0.1128}      & \textbf{0.1734}       & \underline{0.4659}     \\
\cmark                                                                                      & \cmark               & \cmark                   & \xmark                   & \xmark                                    & 0.2126               & 0.2092                & 0.3027                 & 0.1066               & 0.1657                & 0.4559                 \\
\cmark                                                                                      & \cmark               & \cmark                   & \cmark                   & \xmark                                    & 0.1796               & 0.1773                & 0.2748                 & 0.0891               & 0.1411                & 0.4360                 \\
\bottomrule
\end{tabular}
}
\vspace{-2em}
\end{table}

\looseness -1 Therefore, we chose embedding dimensions of 512 and 1024 for the subsequent experiments. However, we will present results with the embedding dimension of 1024, as it consistently yielded better performance for our implementation on ML-20M and MSD datasets.

\noindent \textbf{Item Biases and Regularization}. \, The numerical results of various combinations of these features are present in Table \ref{table:bpr-optimizers}. We can see that models without biases outperform those with item biases in ML-20M across all metrics. Regarding regularization variants, however, the situation is not that apparent. In this dataset, the model incorporating all regularization factors from the original paper \cite{BPR} outperforms others, followed closely by the model using two separate regularization constants for user and item embeddings. This finding suggests that distinct regularization lambdas for users, positive and negative items, are crucial for specific datasets.

\looseness -1 In contrast, the MSD dataset presents different results. Here, the model with item biases performs best overall. However, the effectiveness of item biases varies depending on the regularization factors used. Specifically, biased models surpass unbiased counterparts within two types of regularization approaches: shared and user/item regularization factors. Nevertheless, the difference between biased and unbiased models with shared regularization is statistically insignificant in NDCG@100, determined by a paired t-test with Bonferroni correction ($p < 0.05$) \cite{armstrong2014use}. These findings indicate that the optimal regularization approach and the presence of item biases for BPR depend on the dataset, necessitating careful tuning.

Furthermore, it is notable that the performance of models without regularization is on par with regularized counterparts. In certain instances, these non-regularized models surpassed their regularized counterparts, particularly those utilizing a single regularization factor for user and item embeddings. This phenomenon might be attributed to the effective regularization inherent in the BPR model through the negative sampling algorithm alone.

\begin{table}[t]
\caption{Evaluation of the BPR model in terms of NDCG@K (N@K) and Recall@K (R@K) using various subsets of optimizer and negative sampling features. For ML-20M, the models were trained without item biases and with three separate regularization factors. For MSD, the models were trained without item biases and shared regularization. The best value is bolded, and the second-best is underlined.}
\label{table:bpr-optim-sampler}
\resizebox{\linewidth}{!}{
\begin{tabular}{cccc|cc|cccccc}
\toprule
\multicolumn{6}{c|}{\textbf{Features}}                                                                                                                                                                 & \multirow{7}{*}{N@5} & \multirow{7}{*}{N@10} & \multirow{7}{*}{N@100} & \multirow{7}{*}{R@5} & \multirow{7}{*}{R@10} & \multirow{7}{*}{R@100} \\ \cline{1-6}
\multicolumn{4}{c|}{Optimizer}                                                                                       & \multicolumn{2}{c|}{\begin{tabular}[c]{@{}c@{}}Negative\\Sampling\end{tabular}} & \multicolumn{1}{c}{} & \multicolumn{1}{c}{}  &                        &                      &                       &                        \\ \cline{1-6}
\R{SGD}    & \R{\begin{tabular}[c]{@{}c@{}}Momentum\\SGD\end{tabular}} & \R{RMSProp} & \multicolumn{1}{l|}{\R{Adam}} & \multicolumn{1}{c}{\R{Uniform}} & \multicolumn{1}{c|}{\R{Adaptive}}             & \multicolumn{1}{c}{} & \multicolumn{1}{c}{}  &                        &                      &                       &                        \\ \midrule
\multicolumn{12}{c}{ML-20M}                                                                                                                                                                                                                                                                                                                            \\ \midrule
\cmark     & \xmark                                                    & \xmark      & \xmark                        & \cmark                          & \xmark                                        & 0.2929               & 0.2879                & 0.3883                 & 0.1398               & 0.2171                & \underline{0.6150}     \\
\cmark     & \xmark                                                    & \xmark      & \xmark                        & \xmark                          & \cmark                                        & \underline{0.3085}   & \underline{0.3020}    & \textbf{0.4012}        & \underline{0.1488}   & \textbf{0.2299}       & \textbf{0.6258}        \\
\xmark     & \cmark                                                    & \xmark      & \xmark                        & \cmark                          & \xmark                                        & 0.2669               & 0.2638                & 0.3655                 & 0.1275               & 0.2017                & 0.5939                 \\
\xmark     & \xmark                                                    & \cmark      & \xmark                        & \cmark                          & \xmark                                        & 0.1379               & 0.1393                & 0.2209                 & 0.0641               & 0.1086                & 0.4030                 \\
\xmark     & \xmark                                                    & \xmark      & \cmark                        & \cmark                          & \xmark                                        & 0.2853               & 0.2811                & 0.3818                 & 0.1384               & 0.2157                & 0.6090                 \\
\xmark     & \xmark                                                    & \xmark      & \cmark                        & \xmark                          & \cmark                                        & \textbf{0.3137}      & \textbf{0.3042}       & \underline{0.3986}     & \textbf{0.1510}      & \underline{0.2281}    & 0.6119                 \\ \midrule
\multicolumn{12}{c}{MSD}                                                                                                                                                                                                                                                                                                                               \\ \midrule
\cmark     & \xmark                                                    & \xmark      & \xmark                        & \cmark                          & \xmark                                        & \underline{0.2249}   & \underline{0.2196}    & \underline{0.3132}     & \underline{0.1128}   & \underline{0.1734}    & \underline{0.4659}     \\
\cmark     & \xmark                                                    & \xmark      & \xmark                        & \xmark                          & \cmark                                        & \textbf{0.2475}      & \textbf{0.2395}       & \textbf{0.3289}        & \textbf{0.1253}      & \textbf{0.1881}       & \textbf{0.4730}        \\
\xmark     & \cmark                                                    & \xmark      & \xmark                        & \cmark                          & \xmark                                        & 0.1516               & 0.1518                & 0.2441                 & 0.0771               & 0.1232                & 0.3966                 \\
\xmark     & \xmark                                                    & \cmark      & \xmark                        & \cmark                          & \xmark                                        & 0.0318               & 0.0329                & 0.0627                 & 0.0160               & 0.0274                & 0.1161                 \\
\xmark     & \xmark                                                    & \xmark      & \cmark                        & \cmark                          & \xmark                                        & 0.2014               & 0.2000                & 0.2950                 & 0.1052               & 0.1625                & 0.4484                 \\
\xmark     & \xmark                                                    & \xmark      & \cmark                        & \xmark                          & \cmark                                        & 0.2118               & 0.2088                & 0.3029                 & 0.1106               & 0.1687                & 0.4537                 \\
\bottomrule
\end{tabular}
}
\vspace{-1em}
\end{table}

\noindent \textbf{Optimizers}. \, Table \ref{table:bpr-optim-sampler} provides an overview of the performance exhibited by various optimizer algorithms and negative samplers. In this section, we focus on optimizers utilizing uniform-based negative sampling exclusively. Additionally, although the model with item biases demonstrates the best performance on MSD, we picked the model without them for MSD in this section due to the challenges negative sampling algorithms face with biased models, discussed later in this section.

The table shows that standard SGD demonstrates commendable performance across both datasets, surpassing other optimization algorithms across all metrics. On the other hand, RMSProp shows the poorest performance, particularly on the MSD dataset, where these models primarily minimized regularization constraints.

\begin{figure*}[t]
\centering
\subcaptionbox{MovieLens-20M}{\includegraphics[width=8cm]{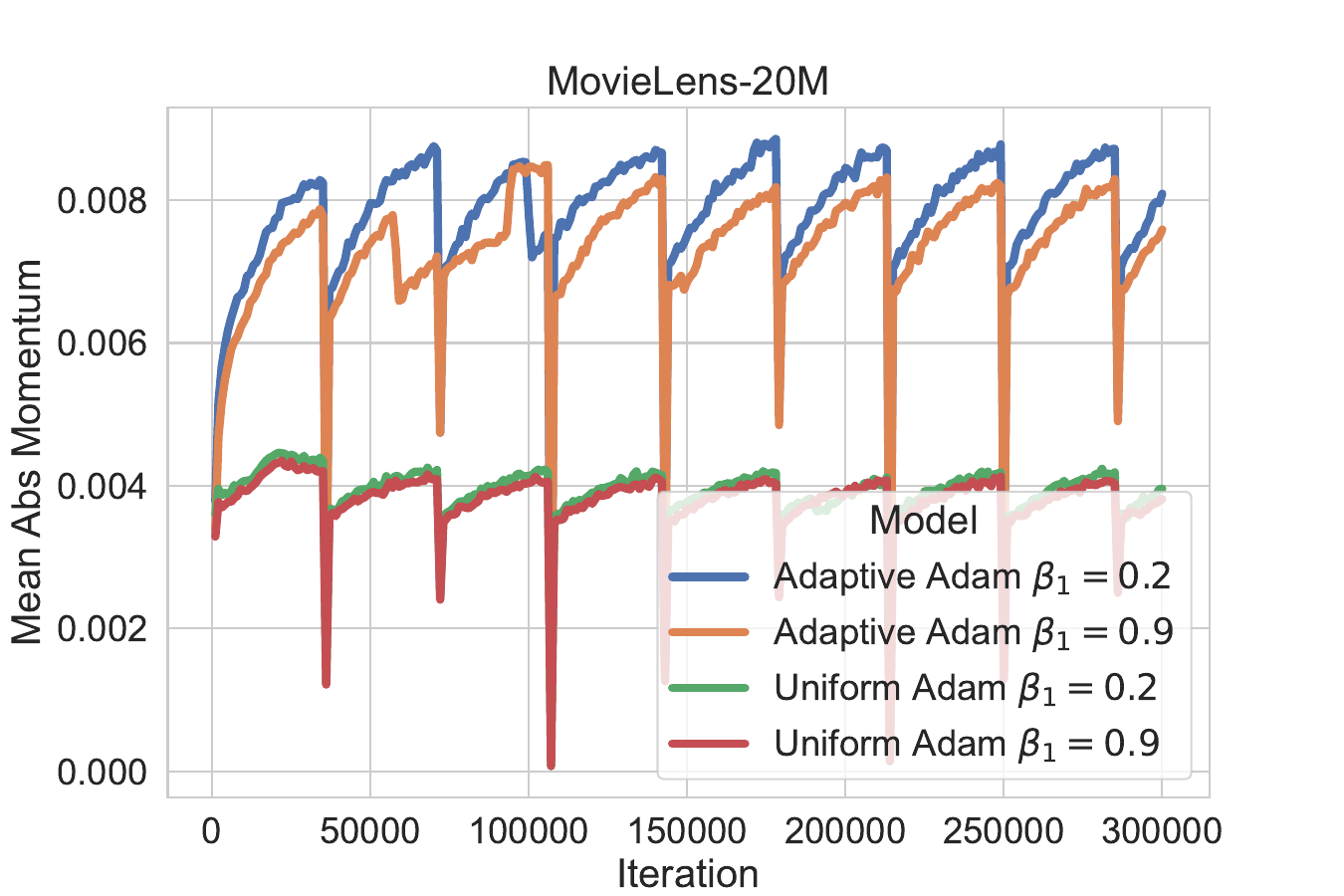}}
\subcaptionbox{Million Song Dataset}{\includegraphics[width=8cm]{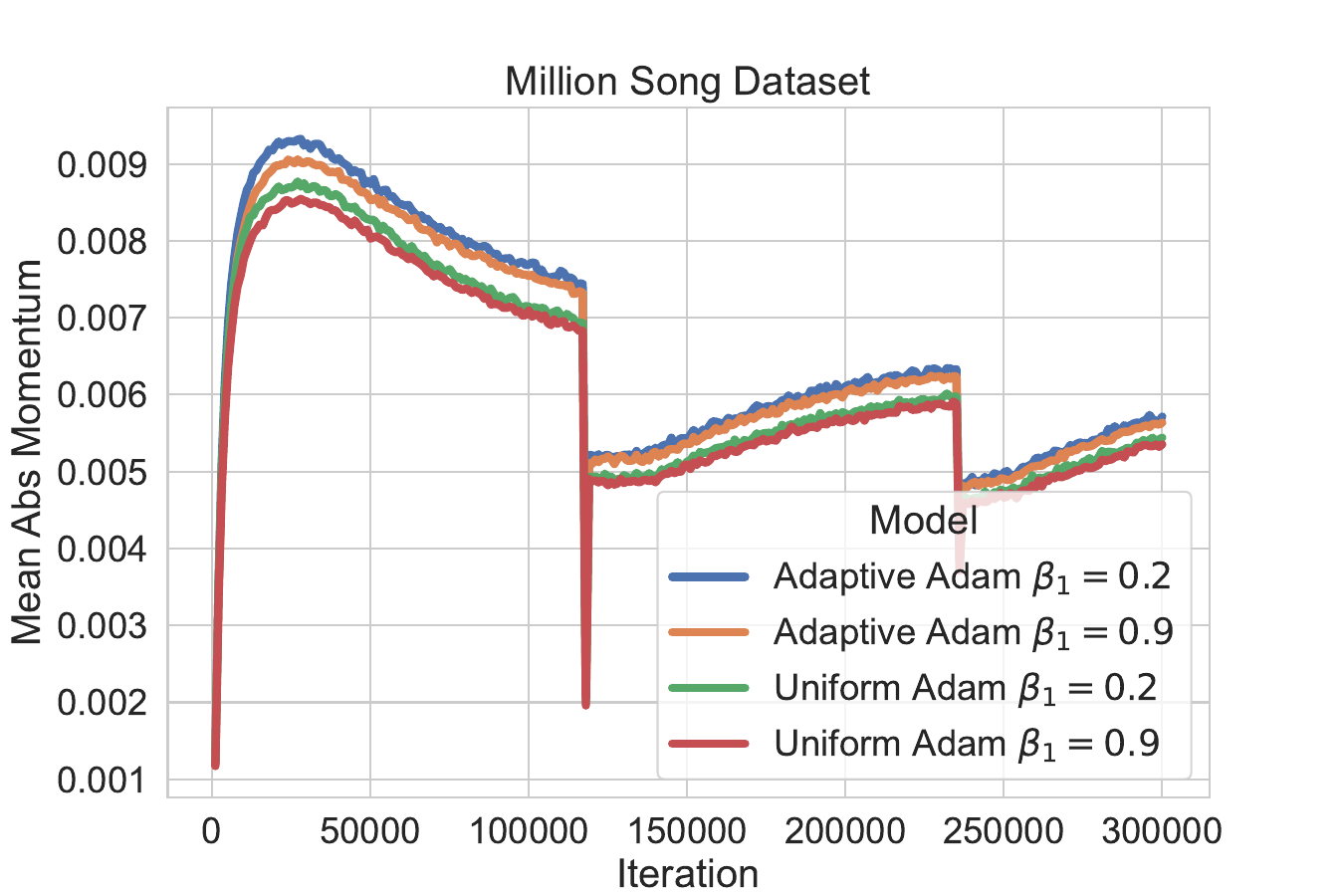}}
\caption{Mean absolute value of the first momentum in the Adam Optimizer with $\beta_1 = 0.2/0.9$ and uniform/adaptive negative sampling for two datasets over the first 300000 training iterations. The values are averaged over every 1000 iterations. Each drop on the graph indicates the beginning of an epoch.}
\label{fig:bpr-adam-momentum}
\vspace{-1em}
\end{figure*}

\looseness -1 Consequently, standard SGD is the preferred choice for the BPR model. However, its crucial drawback must be considered: standard SGD typically requires more epochs than other algorithms. Fortunately, this issue might be mitigated using advanced negative sampling algorithms, which have the potential to significantly reduce the number of epochs required, as we will show later in this section.

Notably, algorithms such as Momentum SGD and Adam require rigorous hyperparameter tuning for optimal performance. We observed that default configurations of these algorithms often yield inferior results. We suppose this phenomenon is associated with uniform-based negative sampling, which tends to generate more simple negatives than strong ones. As highlighted in \cite{bprv2}, this imbalance can result in small gradient magnitudes, slowing model training. This issue especially affects adaptive and momentum-based optimizers as they keep a vector of gradient momenta. We assume that these effects result in poor performance of Recbole and LightFM implementations discussed in Section \ref{section:rq1}.

\looseness -1 Thus, if the momentum primarily comprises gradients with small magnitudes, uniform-based sampling yielding a strong negative can result in smaller weight updates for momentum-based optimizers than standard SGD. Therefore, this dynamic between adaptive optimizers and negative samplers may require more epochs to achieve satisfactory performance. Additionally, we hypothesize that it might lead to poorer performance in other algorithms that employ uniform-based sampling with an adaptive optimizer.

\begin{table*}[t]
\caption{Evaluation results of BPR implementations, baselines, and our BPR implementation in terms of NDCG@K (N@K) and Recall@K (R@K) using two evaluation protocols. For user-based evaluation protocol, the best models for open-source BPR implementations were selected from Figure \ref{fig:bpr-impls-emb-dim}. For our implementation in the user-based evaluation protocol, the best models were selected from Table \ref{table:bpr-optim-sampler}. Values are formatted for convenience$^{14}$.}
\label{table:perf-for-rq3}
\resizebox{\textwidth}{!}{
\begin{tabular}{l|l|cccccc|cccccc}
\toprule
\multicolumn{14}{c}{\textbf{User-based split similar to \cite{rendle2021revisiting}}}                                                                                                                                                                                                                                                                                                                                                \\ \midrule
\multirow{2}{*}{}                                                              & \multirow{2}{*}{Models} & \multicolumn{6}{c|}{ML-20M}                                                                                                                             & \multicolumn{6}{c}{MSD}                                                                                                                                         \\ \cline{3-14}
                                                                               &                         & N@5                   & N@10                  & N@100                   & R@5                     & R@10                     & R@100                    & N@5                      & N@10                     & N@100                    & R@5                      & R@10                     & R@100                    \\
\hline                                                 
\multirow{4}{*}{Baselines}                                                     & ItemPop                 & 0.1298$\dag$          & 0.1275$\dag$          & 0.1906$\dag$            & 0.0580$\dag$            & 0.0912$\dag$             & 0.3298$\dag$             & 0.0360$\dag$             & 0.0349$\dag$             & 0.0582$\dag$             & 0.0176$\dag$             & 0.0271$\dag$             & 0.0986$\dag$             \\
                                                                               & EASE                    & \textbf{0.3376}$\dag$ & \textbf{0.3290}$\dag$ & \textbf{0.4215}$\dag$   & \textbf{0.1612}$\dag$   & \textbf{0.2464}$\dag$    & \underline{0.6371}$\dag$ & \textbf{0.3368}$\dag$    & \textbf{0.3179}$\dag$    & \textbf{0.3907}$\dag$    & \textbf{0.1648}$\dag$    & \textbf{0.2378}$\dag$    & \textbf{0.5096}$\dag$    \\
                                                                               & Mult-DAE                & 0.2949                & 0.2890$\dag$          & 0.3881$\dag$            & 0.1404$\dag$            & 0.2183$\dag$             & 0.6099$\dag$             & 0.2095$\dag$             & 0.1994$\dag$             & 0.2674$\dag$             & 0.1054$\dag$             & 0.1535$\dag$             & 0.3762$\dag$             \\
                                                                               & Mult-VAE                & 0.3071                & \underline{0.3075}    &\underline{0.4158}$\dag$ &\underline{0.1600}$\dag$ & \underline{0.2460}$\dag$ &\textbf{0.6520}$\dag$     & 0.2221$\dag$             & 0.2156$\dag$             & 0.2973$\dag$             & 0.1144$\dag$             & 0.1716$\dag$             & 0.4307$\dag$             \\
\hline                                 
\multirow{6}{*}{\begin{tabular}[c]{@{}l@{}}BPR\\ Implementations\end{tabular}} & Cornac                  & 0.2564$\dag$          & 0.2589$\dag$          & 0.3691$\dag$            & 0.1319$\dag$            & 0.2097$\dag$             & 0.6056$\dag$             & 0.2138$\dag$             & 0.2120$\dag$             & 0.3114$\dag$             & 0.1090$\dag$             & 0.1704$\dag$             & 0.4726                   \\
                                                                               & Implicit                & 0.2071$\dag$          & 0.2052$\dag$          & 0.2880$\dag$            & 0.1093$\dag$            & 0.1673$\dag$             & 0.4745$\dag$             & 0.1749$\dag$             & 0.1737$\dag$             & 0.2567$\dag$             & 0.0929$\dag$             & 0.1425$\dag$             & 0.3898$\dag$             \\
                                                                               & LightFM                 & 0.1295$\dag$          & 0.1262$\dag$          & 0.1843$\dag$            & 0.0544$\dag$            & 0.0869$\dag$             & 0.3087$\dag$             & 0.0362$\dag$             & 0.0349$\dag$             & 0.0575$\dag$             & 0.0175$\dag$             & 0.0268$\dag$             & 0.0964$\dag$             \\
                                                                               & RecBole                 & 0.1242$\dag$          & 0.1213$\dag$          & 0.1785$\dag$            & 0.0513$\dag$            & 0.0820$\dag$             & 0.2973$\dag$             & 0.0004$\dag$             & 0.0009$\dag$             & 0.0117$\dag$             & 0.0002$\dag$             & 0.0008$\dag$             & 0.0328$\dag$             \\
                                                                               & \textbf{Ours (SGD)}     & 0.3085                & 0.3020                & 0.4012                  & 0.1488                  & 0.2299                   & 0.6258                   & \underline{0.2475}       & \underline{0.2395}       & \underline{0.3289}       & \underline{0.1253}       & \underline{0.1881}       & \underline{0.4730}       \\
                                                                               & \textbf{Ours (Adam)}    & \underline{0.3137}    & 0.3042                & 0.3986                  & 0.1510                  & 0.2281                   & 0.6116$\dag$             & 0.2118$\dag$             & 0.2088$\dag$             & 0.3029$\dag$             & 0.1106$\dag$             & 0.1687$\dag$             & 0.4537$\dag$             \\ \midrule
\multicolumn{14}{c}{\textbf{Global temporal split recommended by \cite{hidasi2023evalflaws, dataleakagestudy, meng2020exploring}}}                                                                                                                                                                                                                                                                                                   \\ \midrule
\multirow{2}{*}{}                                                              & \multirow{2}{*}{Models} & \multicolumn{6}{c|}{ML-20M (time-split)}                                                                                                                & \multicolumn{6}{c}{Yelp}                                                                                                                                        \\ \cline{3-14}
                                                                               &                         & N@5                   & N@10                  & N@100                   & R@5                     & R@10                     & R@100                    & N@5                      & N@10                     & N@100                    & R@5                      & R@10                     & R@100                    \\
\hline                                  
\multirow{4}{*}{Baselines}                                                     & ItemPop                 & 0.0939$\dag$          & 0.0855$\dag$          & 0.0990$\dag$            & 0.0169$\dag$            & 0.0274$\dag$             & 0.1312$\dag$             & 0.0034$\dag$             & 0.0041$\dag$             & 0.0105$\dag$             & 0.0030$\dag$             & 0.0055$\dag$             & 0.0306$\dag$             \\
                                                                               & EASE                    & \underline{0.1419}    & \underline{0.1332}    & 0.1698                  & 0.0261                  & 0.0450                   & 0.2496                   & \underline{0.0216}$\dag$ & \underline{0.0250}$\dag$ & 0.0527$\dag$             & \underline{0.0182}$\dag$ & 0.0307$\dag$             & 0.1387$\dag$             \\
                                                                               & Mult-DAE                & 0.1386                & 0.1308                & 0.1691                  & \underline{0.0262}      & 0.0461                   & 0.2455                   & 0.0205$\dag$             & 0.0239$\dag$             & 0.0518$\dag$             & 0.0173$\dag$             & 0.0293$\dag$             & 0.1369$\dag$             \\
                                                                               & Mult-VAE                & 0.1343                & 0.1294                & \underline{0.1722}      & 0.0260                  & \textbf{0.0473}          & \textbf{0.2546}          & 0.0184$\dag$             & 0.0214$\dag$             & 0.0468$\dag$             & 0.0161$\dag$             & 0.0269$\dag$             & 0.1268$\dag$             \\
\hline                            
\multirow{6}{*}{\begin{tabular}[c]{@{}l@{}}BPR\\ Implementations\end{tabular}} & Cornac                  & 0.1315                & 0.1263                & 0.1660                  & 0.0256                  & 0.0445                   & 0.2416$\dag$             & 0.0208$\dag$             & 0.0247$\dag$             & \underline{0.0566}$\dag$ & 0.0176$\dag$             & \underline{0.0310}$\dag$ & \underline{0.1565}$\dag$ \\
                                                                               & Implicit                & 0.1140$\dag$          & 0.1063$\dag$          & 0.1315$\dag$            & 0.0214$\dag$            & 0.0370$\dag$             & 0.191$\dag$              & 0.0118$\dag$             & 0.0140$\dag$             & 0.0348$\dag$             & 0.0101$\dag$             & 0.0178$\dag$             & 0.1004$\dag$             \\
                                                                               & LightFM                 & 0.0983$\dag$          & 0.0900$\dag$          & 0.1038$\dag$            & 0.0169$\dag$            & 0.0285$\dag$             & 0.1375$\dag$             & 0.0022$\dag$             & 0.0025$\dag$             & 0.0030$\dag$             & 0.0015$\dag$             & 0.0026$\dag$             & 0.0052$\dag$             \\
                                                                               & RecBole                 & 0.0033$\dag$          & 0.0031$\dag$          & 0.0061$\dag$            & 0.0004$\dag$            & 0.0008$\dag$             & 0.0095$\dag$             & 0.0001$\dag$             & 0.0001$\dag$             & 0.0005$\dag$             & 0.0001$\dag$             & 0.0002$\dag$             & 0.0018$\dag$             \\
                                                                               & \textbf{Ours (SGD)}     & \textbf{0.1436}       & \textbf{0.1357}       & \textbf{0.1730}         & \textbf{0.0277}         & \underline{0.0471}       & \underline{0.2539}       & \textbf{0.0237}          & \textbf{0.0276}          & \textbf{0.0608}          & \textbf{0.0203}          & \textbf{0.0344}          & \textbf{0.1658}          \\
                                                                               & \textbf{Ours (Adam)}    & 0.1356                & 0.1277                & 0.1637$\dag$            & 0.0248                  & 0.0431                   & 0.2373$\dag$             & 0.0163$\dag$             & 0.0195$\dag$             & 0.0465$\dag$             & 0.0139$\dag$             & 0.0248$\dag$             & 0.1314$\dag$             \\
\bottomrule
\end{tabular}
}
\footnotesize{$^{14}$Within each column, the best value is bolded, the second-best is underlined, and $\dag$ indicates a statistically significant difference ($p < 0.05$) from \textbf{Ours (SGD)} model, determined by a paired t-test with Bonferroni correction \cite{armstrong2014use} for multiple comparisons.}
\end{table*}

\looseness -1 To address this issue, we reduced decay rates for moving averages in adaptive and momentum-based optimizers. This adjustment significantly improved performance, with decay factors in the range of 0.0 to 0.2, yielding better results. Figure \ref{fig:bpr-adam-momentum} illustrates this behavior for the Adam optimizer. We see a slight difference (approximately 2\% on both datasets) between Adam Optimizers with $\beta_1 = 0.2$ and $\beta_1 = 0.9$. It highlights that smaller momentum weights increase gradient magnitudes, improving the training process. However, even with these adjustments, the performance did not surpass that of regular SGD, which does not suffer from this issue.

\noindent \textbf{Negative Sampling}. \, To evaluate the effect of negative sampling, we conducted experiments using the two best-performing optimization algorithms from the previous section, namely SGD and Adam. Table \ref{table:bpr-optim-sampler} presents the results achieved on these optimization algorithms. It is evident from the table that adaptive sampling significantly improves the performance of all algorithms. Additionally, as outlined in \cite{bprv2}, this negative sampling approach samples better negatives with higher gradient magnitudes, allowing models to perform better in fewer epochs. Notably, it reduced the number of epochs required by the SGD optimizer on the ML-20M dataset by 265 steps, resulting in only 85 epochs.

Additionally, Figure \ref{fig:bpr-adam-momentum} illustrates the effect of adaptive negative sampling on the Adam optimizer's first momentum. Obviously, momentum vectors tend to be smaller with uniform-based negative sampling than with adaptive sampling. Specifically, the difference is twice as much on the ML-20M dataset; for MSD, the difference is 5\%. This might partially explain why the model with uniform-based sampling is just 4\% worse than the adaptive-based sampling model.

Surprisingly, when we applied this negative sampling approach to biased models, it consistently resulted in poor performance across all datasets, regardless of other parameters. This phenomenon suggests that adaptive sampling may be ineffective when item biases are present, requiring further investigation in future works.

\subsection{RQ3. Comparison with state-of-the-art models}

We comprehensively evaluated various BPR implementations, comparing them against state-of-the-art baselines to assess their performance across various datasets and evaluation protocols. Table \ref{table:perf-for-rq3} summarizes the results for both user-based splits similar to \cite{rendle2021revisiting} and global temporal splits discussed in \cite{meng2020exploring, hidasi2023evalflaws, dataleakagestudy}. Our implementation of BPR, utilizing both SGD and Adam optimizers, is included for comparison. For both optimizers in the global temporal split, we opt for the model with separated regularization factors $(\lambda_u, \lambda_i, \lambda_j)$, disabled item biases, and adaptive negative sampling, which performed best on the ML-20M dataset according to Section \ref{section:rq2}.

\looseness -1 In the user-based split evaluation, EASE consistently achieved the highest NDCG@K and Recall@K scores across both ML-20M and MSD datasets, demonstrating its robustness in recommending relevant items. Mult-VAE also performed well, particularly in Recall@100, surpassing EASE in some instances. Our SGD-based approach demonstrated competitive performance among the BPR implementations, particularly in the ML-20M dataset, outperforming other BPR implementations in most metrics. Our Adam-based BPR implementation also exhibited robust performance, demonstrating statistically insignificant differences from our SGD-based model across all metrics except Recall@100. Interestingly, our version with SGD optimizer outperformed Mult-VAE on the MSD dataset across all included metrics, while our Adam version occasionally performed worse than Mult-VAE.

The global temporal split evaluation is claimed to be preferred for the experiments \cite{dataleakagestudy, meng2020exploring, hidasi2023evalflaws}. Here, our BPR implementation with SGD optimization outperformed other methods in the ML-20M dataset, achieving the highest NDCG@K scores, although without statistical significance compared to EASE and Mult-VAE. Additionally, in the Yelp dataset, our BPR implementation using SGD optimization excelled, achieving the best results across all metrics with statistical significance. This finding indicates that our BPR implementations are adaptable and can maintain high performance across different datasets and evaluation protocols, highlighting their potential for various recommendation scenarios.

Notably, among the open-source implementations considered, Cornac showed the best performance, aligning with the results obtained in Section \ref{section:rq1}. Other versions from Implicit, LightFM, and RecBole showed decreasing results in this order of performance, with LightFM and RecBole failing to learn anything meaningful.

\section{Conclusions}

In this work, we re-investigated matrix factorization with the BPR optimization criterion. We successfully replicated the BPR model on the Netflix dataset from the original BPR paper. Moreover, we analyzed open-source implementations of the BPR model, uncovering inconsistencies and deviations from the original paper that hindered their performance. Notably, implementation from the Cornac framework achieved the best performances among third-party implementations despite differing from the original model. Also, Elliot is the only open-source framework closely following the BPR paper.

\looseness -1 Furthermore, our investigation into all model's features revealed some intriguing insights. We found that the choice of regularization and negative sampling are pivotal for the model's performance. Surprisingly, we also discovered that the SGD optimizer is a critical factor in achieving good results with the BPR model, as it is tightly linked to the negative sampling process. Also, our findings demonstrate that with proper tuning of these three features, our BPR implementation can achieve performance metrics close to state-of-the-art methods and even surpass some of them on several datasets. For instance, our model statistically significantly outperforms Cornac's results by 8.7\% on ML-20M in NDCG@100 and Mult-VAE on MSD by 10\%. Moreover, our implementation showcases exceptional results on datasets with global temporal split, surpassing even Mult-VAE and EASE. We hope these findings inspire further research into BPR-based extensions and other models that utilize the BPR objective, such as LightGCN, GRU4Rec, and Transformers4Rec.

\begin{acks}
The contribution is an output of a research project implemented in the TBank and the Laboratory for Models and Methods of Computational Pragmatics at the National Research University Higher School of Economics (HSE University).
\end{acks}

\bibliographystyle{ACM-Reference-Format}
\bibliography{references}

\end{document}